\documentclass[structabstract]{aa}
\usepackage{url}
\usepackage{graphicx}
\usepackage{txfonts}

\newcommand{\solm}{M$_{\odot}$}
\newcommand{\unit}[1]{\,\ensuremath{\mathrm{#1}}}

\begin{document}

\title{High-resolution observations of SDSS~J080800.99+483807.7 in the
optical and radio domains}
\subtitle{A possible example of jet-triggered star formation} 
\titlerunning{SDSS~J080800.99+483807.7}

\author{Y. E. Rashed\inst{1,2}
\and J. Zuther\inst{1}
\and A. Eckart\inst{1,3}
\and G. Busch\inst{1}
\and M. Valencia-S.\inst{1}
\and M. Vitale\inst{1,3}
\and S. Britzen\inst{3}
\and T. Muxlow\inst{4}}

\institute{I. Physikalisches Institut, Universität zu K\"oln, Z\"ulpicher Stra\ss e 77, 50937 K\"oln, Germany
\and Department of Astronomy, Faculty of Science, University of Baghdad, 10071 Baghdad - Aljadirya, Iraq\\ \email{yasir@ph1.uni-koeln.de}
\and Max-Planck-Institut f\"ur Radioastronomie, Auf dem H\"ugel 69, 53121 Bonn, Germany
\and MERLIN/VLBI National Facility, Jodrell Bank Observatory, University of Manchester, Macclesfield, Cheshire SK11 9DL, UK}


\abstract{
Double-lobe radio galaxies are ideally suited to investigate the interaction 
of the individual components of the radio structure with the intergalactic medium and 
the interstellar medium of the host galaxy. SDSS J080800.99+483807.7 has been 
serendipitously discovered in MERLIN 18 cm observations to be a double-lobed radio galaxy. 
Because it is an optically faint source, basic information like redshift, linear size, and structure has been 
incomplete until now. Furthermore, there are no spectra of this source available in any databases.}
{The goal of this work is to derive the main physical properties of
SDSS~J080800.99+483807.7 and study the possible interaction between the radio
jets and the interstellar medium of the host galaxy.}
{To achieve this goal, we used optical spectroscopy and radio interferometry.
The radio data were obtained with MERLIN at 18\,cm and the optical data with the Multi-Object Double Spectrograph 
(MODS) at the Large Binocular Telescope (LBT).} 
{The redshift of the galaxy is $z=0.2805\pm 0.0003$, resulting in a linear size
of the observed radio structure of $\sim 26.3\unit{\rm kpc}$. 
The optical line emission as well as the infrared and radio continuum emission suggest a high
star-formation activity. In addition, we estimated the mass of the central black
hole to be $\log \left(M_{\rm BH}/M_\sun \right)\approx 6.9$.} 
{The MODS spectrum and the optical images from Sloan Digital Sky Survey suggest that
SDSS~J080800.99+483807.7 is an elliptical host galaxy.
In combination with the overall radio structure, we argue that the
star formation could be the result of the back-flow along the jet and the interstellar medium of
the host.}


\maketitle

\section{Introduction}
The first observations of radio galaxies have provided very strong evidence
that there are extended jets flowing from the center of some of these sources, spanning 
a few kilo parsec to a few mega parsecs
\citep[e.g.,][ and references therein]{1984ARA&A..22..319B}. 
The jets appear to be associated with the central massive black hole
and the accretion
disk \citep[e.g.,][]{1984RvMP...56..255B, 1977MNRAS.179..433B, 1982MNRAS.199..883B}.
It has been widely speculated about the mechanisms for fueling the central massive
black hole with matter from the host galaxy and the impact of nuclear activity 
on the host (i.e., radiation field, outflows, jets, etc.).
\cite{2012ARA&A..50..455F} reports on observational evidence for feedback
between the radio jet and the interstellar/intergalactic medium, especially in
the radio/kinematic mode in which a significant back-flow of material along the
periphery of the jet or outflow onto the host occurs \citep{2013ApJ...763L..18W,2013ApJ...772..112S}. 
\cite{2013ApJ...763L..18W} use three-dimensional grid-based hydrodynamical simulations to show that ultrafast outflows 
from active galactic nuclei (AGN) result in considerable feedback of energy and momentum into 
the interstellar medium (ISM) of the host galaxy. 
\cite{2013ApJ...772..112S} shows that the AGN-induced pressure, which is caused by jets and/or winds that flow back onto a gas-rich host, 
can lead to pressure-regulated star formation with significantly enhanced star-formation rates.
The back-flow phenomenon has also been discussed, especially in the case of X-shaped radio galaxies (XRGs), 
which are a special type of radio galaxies with
two pairs of lobes \citep[one active, bright pair and a second fainter, more diffuse pair;][]{1984MNRAS.210..929L}.
For a radio galaxy to be classified as an XRG, it needs to have the two pairs of jets strongly misaligned and flowing from 
the center of the source; the minor pair must flow in another direction as the major pair, otherwise 
it would be double-double radio galaxy \citep[DDRG; e.g.,][ and references therein]{2011MNRAS.410..484B}.
One finds that 5\%-10\% of Fanaroff-Riley type II radio galaxies are X-shaped galaxies
\citep{1974MNRAS.167P..31F}. 
There are several competing interpretations of the physical nature of the radio
morphology, e.g., black-hole spin reorientation, plasma back-flows from the
lobes, binary black holes, and jet interstellar-medium interaction
\citep[cf.][for a review]{2012RAA....12..127G}. 
The nuclei of XRGs, which presumably represent the transition population
between FR~I and FR~II radio sources \citep[e.g.,][]{2010MNRAS.408.1103L}, are
valuable probes of the interaction between the black hole, its jet, and the host
\citep[e.g.,][]{2011ApJ...733...58H, 2011ApJ...728...29W}. 

In our MERLIN 18\,cm data set of 4C~48.21 we have serendipitously 
identified SDSS~J080800.99+483807.7 (in the following J0808) as a rather compact double-lobed radio galaxy. 
Such objects are particularly interesting since they could be small due to extreme
foreshortening or an actual small size. In case of an extreme
foreshortening, one would expect the radio jet to be pointed towards the observer
under a small angle with respect to the line of sight, such that the central nucleus appears
to be bright due to relativistic beaming. 
In the latter case, one of the lobes is also expected to be brighter and possibly
cover the central source due to its extent. However, J0808 appears as a
symmetric double-lobed 
source with a nuclear component that is very weak with
respect to the lobes and somewhat more extended than compact. 
Hence, we can assume that its small overall apparent 
size is due to a small physical size. 
Especially for physically small sources, a
back-flow towards the center along the jets is expected, resulting in an
interaction at the nucleus. 
The possibility of back-flow interactions has also been 
reported \citep[cf.][]{2011ApJ...733...58H} for X-shaped radio sources. 
In particular, we complement the radio information with optical
spectroscopy of the nuclear region of the galaxy and investigate the
possibility of triggered star formation.
A detailed description of the observations and the J0808 data reduction in this 
context is given in section \ref{sec:obs_red}.
In Section \ref{sec:results} we discuss the results:
the optical spectrum of the source (Sect. \ref{sec:em_lines}),
the radio structure and its linear size (Sect. \ref{sec:lin_size}),
the radio loudness of J0808 (Sect. \ref{sec:radio}),
the central black hole  (Sect. \ref{sec:bh}),
and finally in Sect. \ref{sec:star_form}, the star formation in the host of J0808.
A discussion follows in Section \ref{sec:discussion}.

\begin{figure*}
\begin{center}
\includegraphics*[width=12cm]{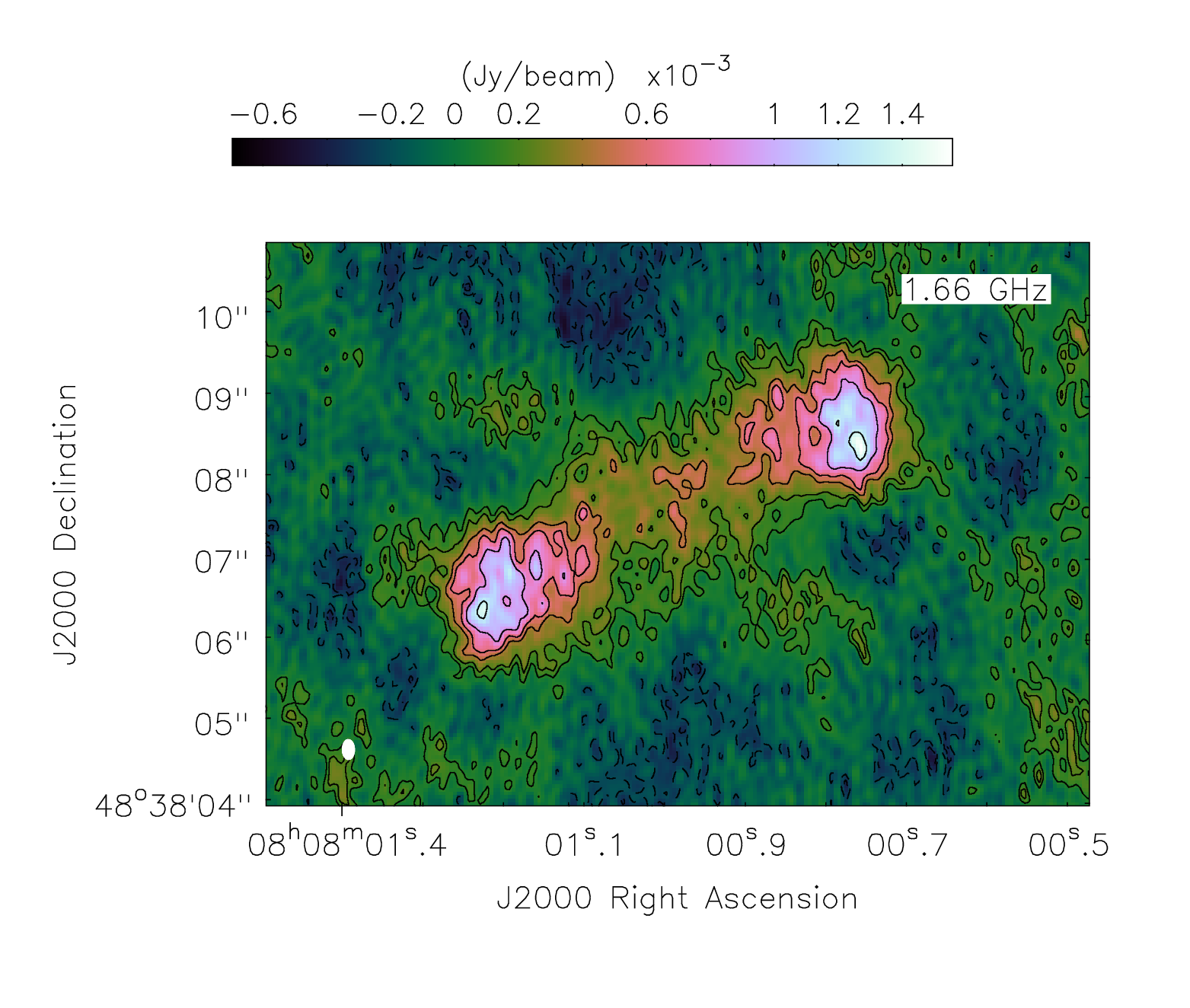}
\includegraphics*[width=12cm]{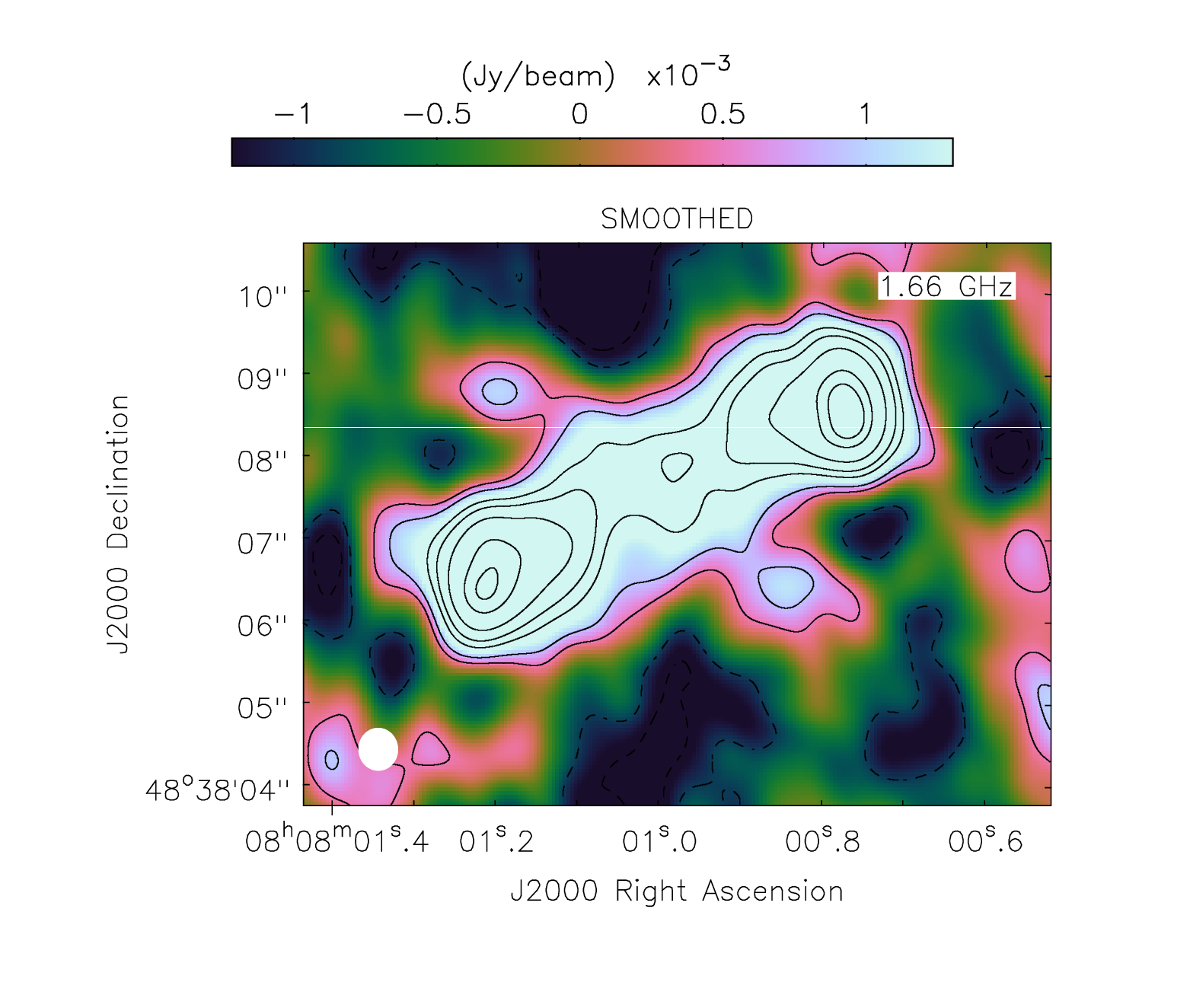}
\caption{18~cm MERLIN image of SDSS J080800.99+483807.7.
The J2000 coordinates of the central source are $\alpha=$08:08:00.99 and
$\delta=$+48:38:07.73.
{\it Top:} original image at a noise level of about 1.1e-4 Jy/beam.
{\it Bottom:} 10 pix smoothed image at a noise level of about 4.5e-4 Jy/beam.
The contour levels are each at 
-3, -2, 1, 2, 4, 5.5, 8, 12, 14 times the noise level. 
}
\label{ra_im}
\end{center}
\end{figure*}

\begin{figure*}
\begin{center}
\includegraphics*[width=0.9\linewidth]{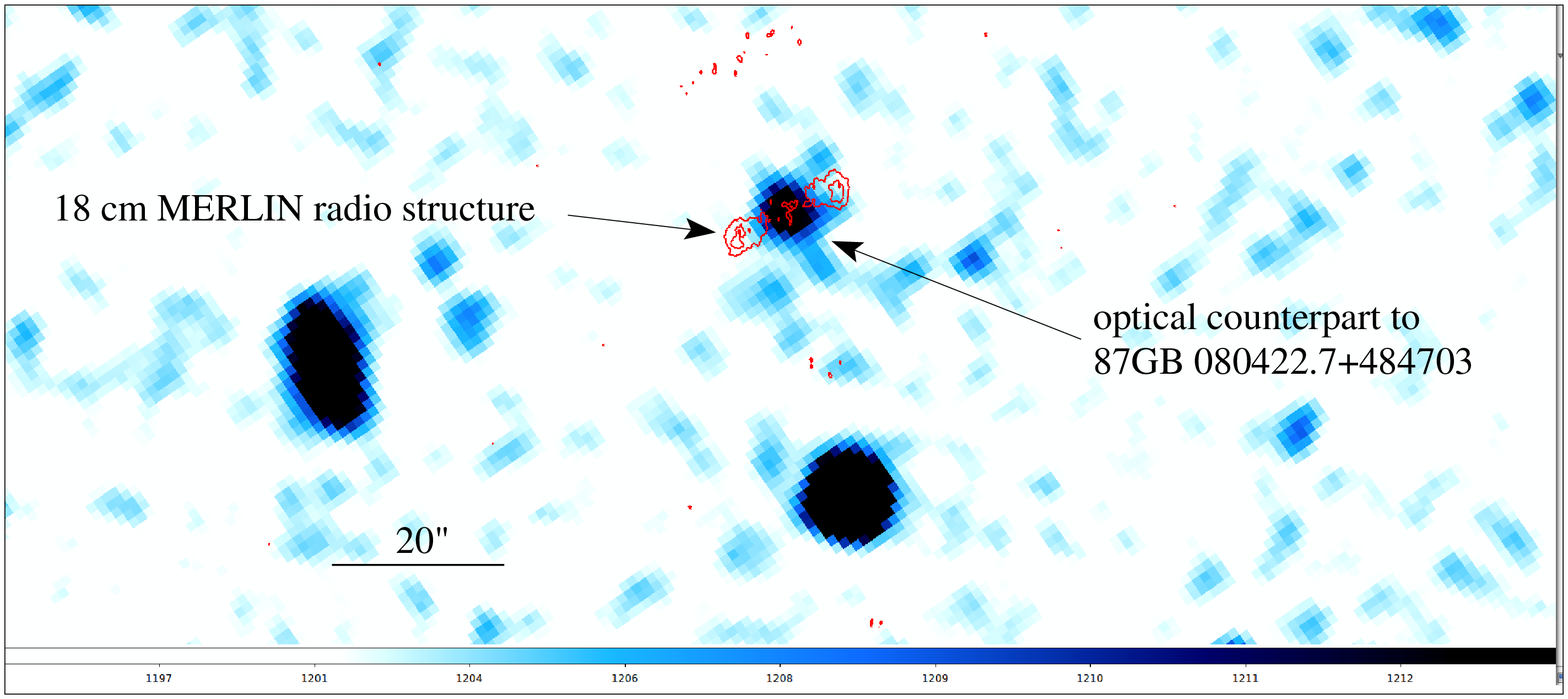}
\caption[]{SDSS image of J0808, overlaid with MERLIN 18\,cm contours in red.}
\label{optical_sdss}
\end{center}
\end{figure*}

\section{Observation and Data Reduction}
\label{sec:obs_red}
In this section, we describe the radio and spectroscopic
observations, the data reduction, and the complementary data from public
archives used in our analysis.

\subsection{MERLIN 18\,cm observation}
The Multi-Element Radio Linked Interferometer Network (MERLIN) is an
interferometer array of radio telescopes distributed across England with
baselines up to 217~km length. 
The 18~cm observations were carried out in May 2005.
The resulting angular resolution is about $0\farcs 25\times 0\farcs 15$ with a
position angle of 1.5$^\circ$.
The total integration time on J0808
 was about 9.2 hours and the noise in the
image is about 0.1~mJy~beam$^{-1}$. 
The central radio component is located within 0\farcs 1 of the
position of the optical source J0808.
Fig. \ref{ra_im} presents the image of the radio source.
Details of the radio observation and data reduction are given in
\cite{2012A&A...543A..57Z}.
An extended structure at the central source component of J0808 gives some evidence for 
a possible back-flow towards the center that may result in an interaction 
with the ISM at the nucleus.
In fact, J0808 shows extended emission along a position angle that is by at least
50$^\circ$ different from that of the prominent double-lobe structure.
Signatures of the peculiar shape are prominent on contour lines as high as
at least five times the noise level in the original and the smoothed version of the 
radio image shown in Fig. \ref{ra_im}.
This is a signature well known for XRGs or in general objects 
that are subject to a significant 
back-flow along the radio jet onto the host galaxy 
\citep{2013ApJ...763L..18W,2013ApJ...772..112S}.

\subsection{Multi-Object Double Spectrograph observation} 
The Multi-Object Double Spectrograph (MODS) is mounted at the Large
Binocular Telescope (LBT), located on Mount Graham in the Pinaleno Mountains,
USA. It operates between 3200 and 10500$\AA$ at a
nominal spectral resolution of $\lambda/\delta\lambda\approx2000$ (i.e., $\sim150\ \mathrm{km}\ \mathrm{s}^{-1}$). 
Its field of view is about $6\times6$ arc-minutes$^2$. 
Furthermore, imaging, long-slit, and multi-object
spectroscopy can be used in this instrument, with a slit aperture of 0\farcs 6. The MODS1 spectrograph
had first light in September 2010 and began its science operation in September
2011. The MODS charge-coupled device (CCD) detector is an e2vCCD231-68 $8k \times 3k$ with a 15 $\mu$m
pixel pitch. For the grating spectroscopy, the full spectral range was split into a blue and
red channel, which are 3200-6000 $\AA$ (blue) and
5000-10500 $\AA$ (red) \citep{2010SPIE.7735E...9P,2012SPIE.8446E..0GP}. 
For a slit width of 0\farcs 6 and a redshift of $z\approx 0.28$, we probe the
central 2.5~kpc region of the host.

\begin{figure*}
\begin{center}
\includegraphics*[width=\linewidth]{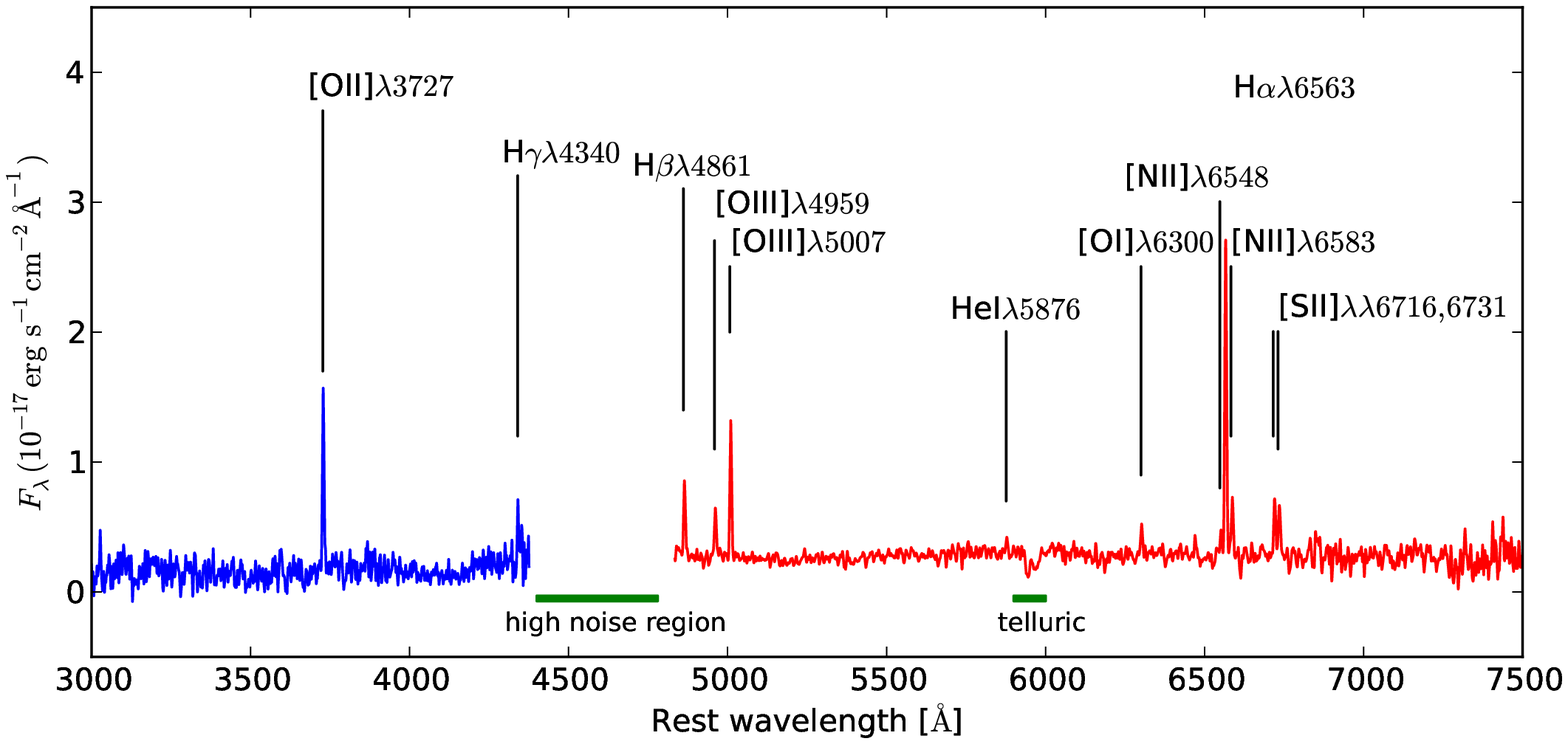}
\end{center}
\caption{Optical spectrum of J0808. The blue and red spectra represent the data from the respective channel. 
A gap between the two channels that has been excluded because of
high noise and a region that shows telluric features are marked.}  
\label{op_sp}
\end{figure*}

For data reduction, we used the modsCCDRed software provided for MODS1, built
in Python.
This software package allows us to perform bias 
correction, flat fielding, and other standard steps of data reduction.

As a first step, we prepared  normalized spectral flat field images for 
each channel (blue and red) through the following
\footnote{The user manual can be found at
\url{http://www.astronomy.ohio-state.edu/MODS/Manuals/MODSCCDRed.pdf}. }:
\begin{itemize}
\item corrected the bias and trimmed the flat fields images;
\item median combination of the bias-corrected flats;
\item repaired the bad columns using the bad pixel lists for the detector;
\item eliminated the color term to produce a normalized ``pixel flat''.
\end{itemize}

In a second step, we applied these two normalized spectral flat fielding images 
to all raw science and calibration frames. In a third step, we performed the  
wavelength calibration for science frames using the lamp frames provided by the 
instrument for different elements (argon, neon, xenon and krypton). 
Here, we used the \textsc{Iraf} reduction package to identify the lamp lines and
then transformed the science frames accordingly.
In addition, we tested the calibration using the OH-skyline atlas
\citep{1996PASP..108..277O, 1997PASP..109..614O}. We found both ways of 
calibration to be in good agreement. Then, in a final step we extracted 
one-dimensional spectra from the two-dimensional images and flux-calibrated 
them using the standard star G191B2B.

\subsection{The Sloan Digital Sky Survey}
The Sloan Digital Sky Survey (SDSS) is a sensitive optical imaging and
spectroscopic public survey of about $10^4$ square degrees of the northern sky
\citep{2000AJ....120.1579Y}.
It provides images and photometric parameters in five bands ($u$, $g$, $r$, $i$, 
and $z$) at an average seeing of 1\farcs 5 and down to a limiting
magnitude of $\sim 22.2$ in $r$ band.
The $R\sim 2000$ spectroscopy covers a wavelength range from 3800-9200 \AA.
SDSS began its operation in May 2000.  
We used the seventh data release of the SDSS \citep{2009ApJS..182..543A}. 
Magnitudes for J0808 in the five SDSS bands are given in Table. \ref{Table_mag}.
In Fig. \ref{optical_sdss}, we show the $i$-band SDSS optical image of the
galaxy. 

\subsection{The Wide-field Infrared Survey Explorer}
The Wide-field Infrared Survey Explorer (WISE) was a NASA 
infrared-wavelength astronomical space telescope, which was operated 
between  December 2009 to February 2011 \citep{2010AJ....140.1868W}. 
It performed an all-sky astronomical survey with images 
in wavelength bands at 3.4~$\mu$m, 4.6~$\mu$m, 12~$\mu$m, and 22~$\mu$m, using a
40~cm  diameter infrared telescope in Earth orbit. 
In the all-sky data release from March 14, 2012, the source J0808 is detected at
all bands, except at 22~$\mu$m. The flux densities are listed in
Table~\ref{Table_mag}. 
The uncertainty of the 12~$\mu$m flux density was estimated from the
corresponding image provided by the data release. 

\begin{table}
\begin{center}
\caption{Photometric measurements of J0808
 in the SDSS and WISE photometric systems.
}

\label{Table_mag} 
\begin{tabular}{c c c}       
\hline\hline
Survey & Band& Magnitude \\
\hline
SDSS  & $u$& $23.68 \pm 1.35$\\
SDSS  & $g$& $24.05 \pm 0.91$\\
SDSS  & $r$& $21.9 \pm 0.20$\\
SDSS  & $i$& $20.52 \pm 0.10$\\
SDSS  & $z$& $19.70 \pm 0.15$\\
WISE  & 3.4 $\mu$m & $15.61 \pm 0.06$\\
WISE  & 4.6 $\mu$m & $15.51 \pm 0.15$\\
WISE  & 12 $\mu$m & $12.30 \pm 2.50$\\
\hline                                            
\end{tabular}
\end{center}
\end{table}

\section{Results}
\label{sec:results}

\subsection {Measurement of emission-lines}
\label{sec:em_lines}
 
We present the reduced optical spectrum in Fig. \ref{op_sp}. 
Prominent emission lines have been fitted with the spectral analysis tool
Specview assuming Gaussian shape.  
Positions, resulting redshifts, flux density, and full-width at half maximum 
(FWHM) of the emission lines are listed in Table~\ref{FWHM}.\\
The emission-line diagnostic diagram delivers information about the
dominant ionization mechanism of the gas in the studied region. 
Studies of different samples of narrow emission-line galaxies 
have also shown some trends in the diagrams that are associated with stellar mass, 
metallicity and possibly morphology of the host. 
No reddening corrections has to be applied to calculate line ratios in these diagrams 
since the emission lines are close to each other in wavelength:

\begin{itemize}
\item \ [\ion{N}{II}]/H$\alpha$ vs. [\ion{O}{III}]/H$\beta$
\citep{1987ApJS...63..295V,2001ApJ...556..121K,2003MNRAS.346.1055K,
2003AAS...20311901K, 2004MNRAS.350..396L,
2010A&A...509A..53L,2011ApJ...736..104J}.  

\item \ [\ion{S}{ii}]/H$\alpha$ vs. [\ion{O}{iii}]/H$\beta$
\citep{1987ApJS...63..295V,2001ApJ...556..121K,2004MNRAS.350..396L,
2006MNRAS.372..961K,
2010A&A...509A..53L,2011ApJ...736..104J}.  

\item \ [\ion{O}{i}]/H$\alpha$ vs. [\ion{O}{iii}]/H$\beta$
\citep{2001ApJ...556..121K, 2006MNRAS.372..961K}. 
\end{itemize}

The measured  ratios between emission lines are:
\begin{itemize}
\item $\log \frac{[\ion{O}{iii}]\ \lambda5007}{\mathrm{H}\beta}=0.337\pm0.006$
\item $\log \frac{[\ion{N}{ii}]\ \lambda6583}{\mathrm{H}\alpha}=-0.822\pm0.005$
\item $\log \frac{[\ion{S}{ii}]\ \lambda6716, \lambda6731}{\mathrm{H}\alpha}=-0.448\pm0.006$
\item $\log \frac{[\ion{O}{i}]\ \lambda6300}{\mathrm{H}\alpha}=-1.090\pm0.010$
\end{itemize}

For our target source only 
an estimate of a photometric redshift of $z\sim0.7$ was available from the SDSS.
To derive the spectroscopic redshift, we first reduced the 
data as outlined in the previous section. The optical spectrum is shown in
Fig. \ref{op_sp}.
For the emission lines evident in Fig. \ref{op_sp} and listed in Table
\ref{FWHM}, we used  
the equation
\begin{equation}
     z=\frac{\lambda_\mathrm{obs}}{\lambda_\mathrm{rest}} -1
\end{equation}
and calculated the redshift for each line as shown in Table \ref{FWHM}. 
The mean redshift and its standard deviation are $z=0.2805\pm0.0003$. 
Combining the redshift with
the cosmology constants $H_0=70\ \mathrm{km} \ \mathrm{s}^{-1} \
\mathrm{Mpc}^{-1}$,  
$\Omega_{\mathrm m}=0.3$, and $\Omega_\Lambda=0.7$ \citep[][which we use
throughout the paper]{2003ApJS..148..175S}, 
we calculated the luminosity distance $D_\mathrm{L}$ 
\citep{1999astro.ph..5116H} to be $1437.2\ \mathrm{Mpc}$.
This corresponds to a linear-size scale conversion factor of 6.98 kpc per arcsecond.

\begin{figure*}
\includegraphics[width=\linewidth]{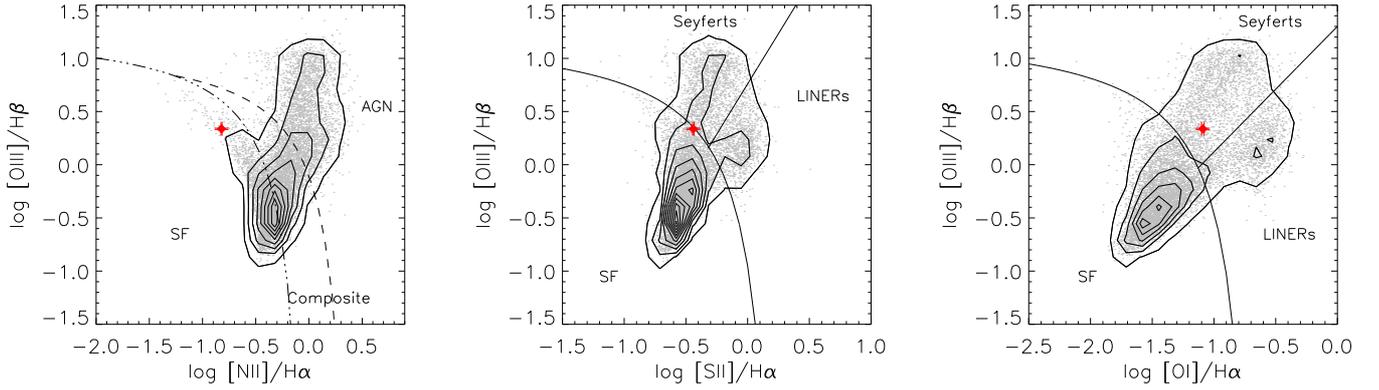}
\caption{
Diagnostic diagrams [\ion{N}{ii}/H$\alpha$ vs.
[\ion{O}{iii}]/H$\beta$, [\ion{S}{ii}]/H$\alpha$ vs. [\ion{O}{iii}]/H$\beta$ and
[\ion{O}{i}]/H$\alpha$ vs. [\ion{O}{iii}]/H$\beta$ for the radio emitters taken from
\cite{2012A&A...546A..17V}. The contours show galaxy densities in steps of 80 galaxies
per each additional contour. The position of the source analyzed here is shown as a red dot.
}
\label{vitale}
\end{figure*}

As we see in the diagnostic diagrams (Fig.~\ref{vitale}), J0808 is located in the region
characteristic for host galaxies with high star-formation rates 
([\ion{N}{II}]-based diagram)
and a tendency to have contributions to line emission 
from an active nucleus ([\ion{O}{I}]-based diagram).
J0808 is located closer to the region of star formation and low excitation galaxies, 
as compared to high excitation galaxies by \cite{2009AN....330..210K}.
In particular, the low [\ion{N}{II}]/H$\alpha$ also suggests a low metallicity.
The diagnostic diagrams in Fig.12 by 
\cite{2012A&A...546A..17V} suggest a thermal rather than a nonthermal origin of the
optical emission lines.
From the optical images it is difficult to provide a  host galaxy classification.
The host visible in the SDSS images is in agreement either with an elliptical 
or a bulge of a low-luminosity disk system.
The low metallicity then suggests either an early evolutionary stage or an
overall  low mass of the host galaxy \citep{2003MNRAS.346.1055K,2013arXiv1304.2776V}.
Interpolating the $z$-band and 3.4~$\mu$m flux densities, we get a K-band magnitude of about
K$\sim$16.9 and a flux density of S$_K$=1.09$\times$10$^{-4}$~Jy. 
We can estimate the old stellar mass via
\begin{equation}
     M_{*,{\rm old}}[M_{\odot}]=2.6 \times 10^8 D^2[{\rm Mpc}] S_K[{\rm Jy}]\ 
\end{equation}
\citep{1988ApJ...327..671T}.
This is equivalent to a population K-band M/L ratio of about 23 and applying 
the K-band luminosity to flux density relation by \cite{1994ApJ...425...72K}. 
Using a distance of 1437.2~Mpc, we then find 
a stellar mass of about 6$\times$10$^{10}$\solm.
This is one to two orders of magnitudes less than the mass of the Milky Way 
but about an order of magnitude more than the upper limit 
of 2$\times10^9$\solm ~derived for 
the stellar mass of the starburst galaxy M82
\citep{2012ApJ...757...24G}.
M82 is known to have a strong, nuclear starburst driven wind over which 
a major portion of the newly formed metals can be lost.
Hence, J0808 could be a low-mass low-metallicity elliptical.

\subsection{Linear size of the radio source}
\label{sec:lin_size}

Early investigations of the relation between the median linear size of the galaxy 
 $D_\mathrm{med}$ as a function of redshift $z$ \citep{1985MNRAS.217..179E} 
resulted in 
$D_\mathrm{med} \propto (1+z)^{-1.1\pm 0.5}$ 
for $\Omega_{0}=0$,  
and
$D_\mathrm{med} \propto (1+z)^{-1.45\pm 0.4}$  
for $\Omega_{0}=1$.
Initially, there was no indication for a correlation between 
luminosity and linear size. 
However, \cite{1987Natur.328..500O}  present 
a strong correlation between $D_\mathrm{med}$ and $z$ 
of $D_\mathrm{med} \propto (1+z)^{-3.3\pm 0.5}$ at $\Omega_{0}=0$. 
Follow-up investigations by \cite{1995ApJ...451...76N} 
could not confirm such a strong correlation and resulted in
an updated relation between the median size and redshift of
$D_\mathrm{med} \propto (1+z)^{-1.2\pm 0.5}$ when $\Omega_{0}=0$ and 
$D_\mathrm{med} \propto (1+z)^{-1.7\pm 0.4}$ when $\Omega_{0}=1$.
In recent years, spectroscopic techniques have helped 
to understand the physical properties of more distant objects that 
appear to be largely unresolved.
Using spectroscopic redshifts allows us to derive the linear size of the
radio source host galaxies.

\begin{figure}
\centering
\includegraphics*[width=\columnwidth]{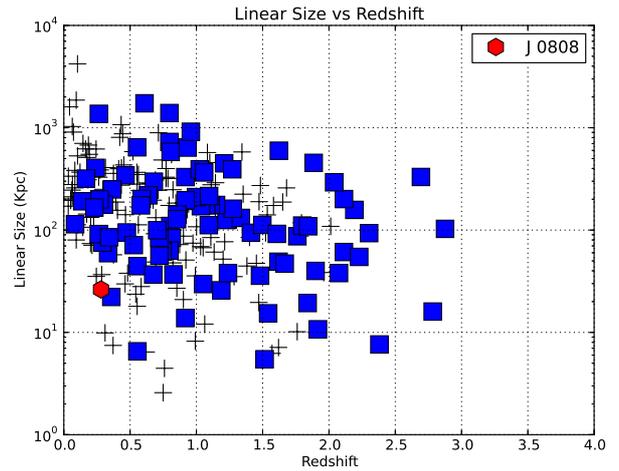}
\caption{Linear size versus redshift.  Our target source is
show by the red hexagon, the plus signs represent the location of the 
3CRR sources, the square signs represent the 7CRS sources \citep{2011MNRAS.418.1138W}.}
\label{fig.linear}
\end{figure}

We derive a linear source size via
\begin{equation}
    l=\vartheta \times D_\mathrm{A}=
   \frac{\vartheta \times D_\mathrm{L}}{(1+z)^{2}}\ .
\end{equation}
Here 
$D_\mathrm{A}$ is the angular diameter distance and
$\vartheta=6\farcs2$
is the angular size of our source J0808  as derived 
from the radio image shown in Fig. \ref{ra_im}. 
This results in a linear size of $l=26.29\ \mathrm{kpc}$, which 
can be compared to the size values found for other sources, 
as given in, e.g., \cite{2011MNRAS.418.1138W}. 
The comparison is shown in Fig. \ref{fig.linear}, and 
we see that the J0808 is a relatively compact source 
compared to other galaxies at similar redshift. 

\begin{figure}
\begin{center}
\includegraphics*[width=\columnwidth]{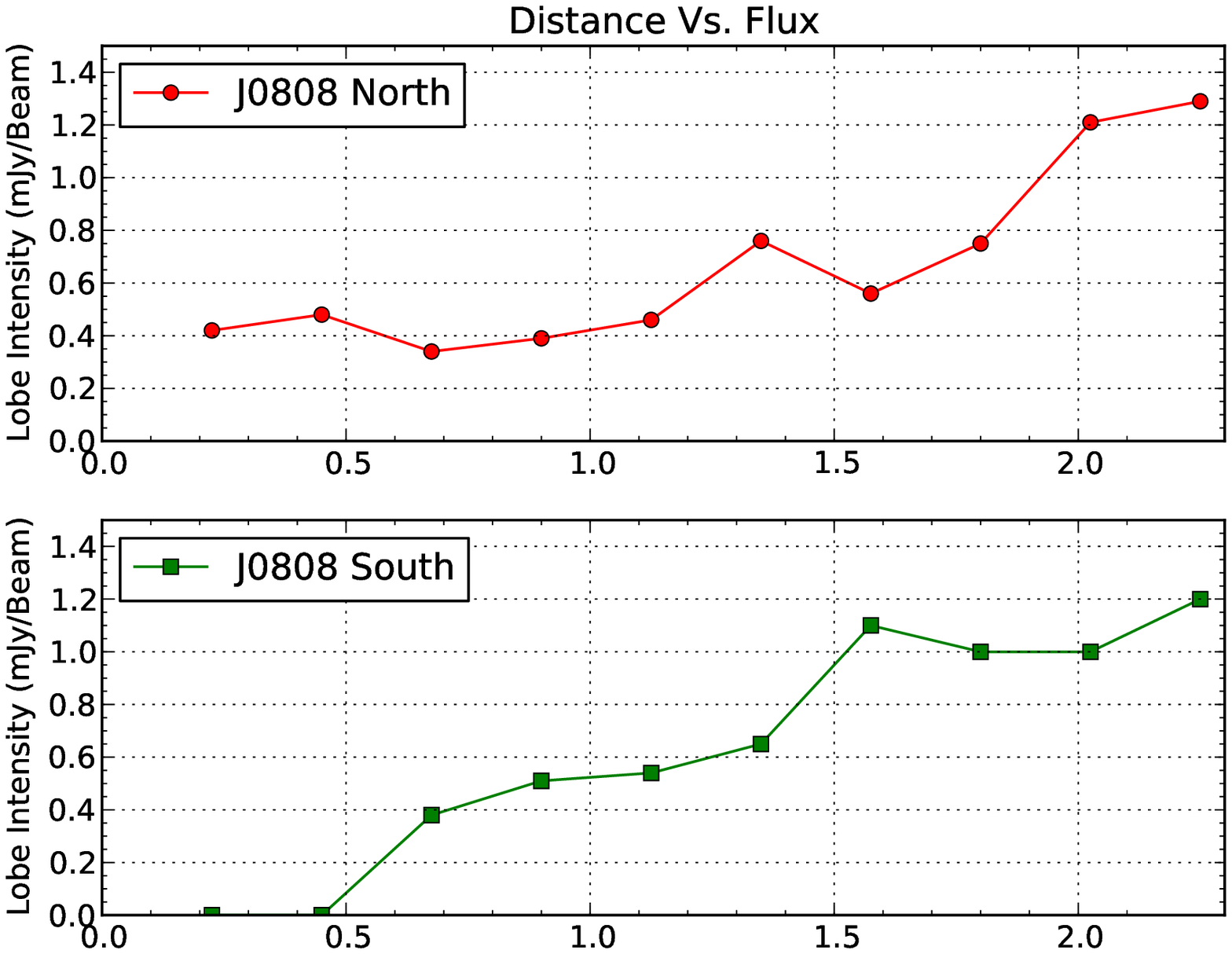}\\
\includegraphics*[width=\columnwidth]{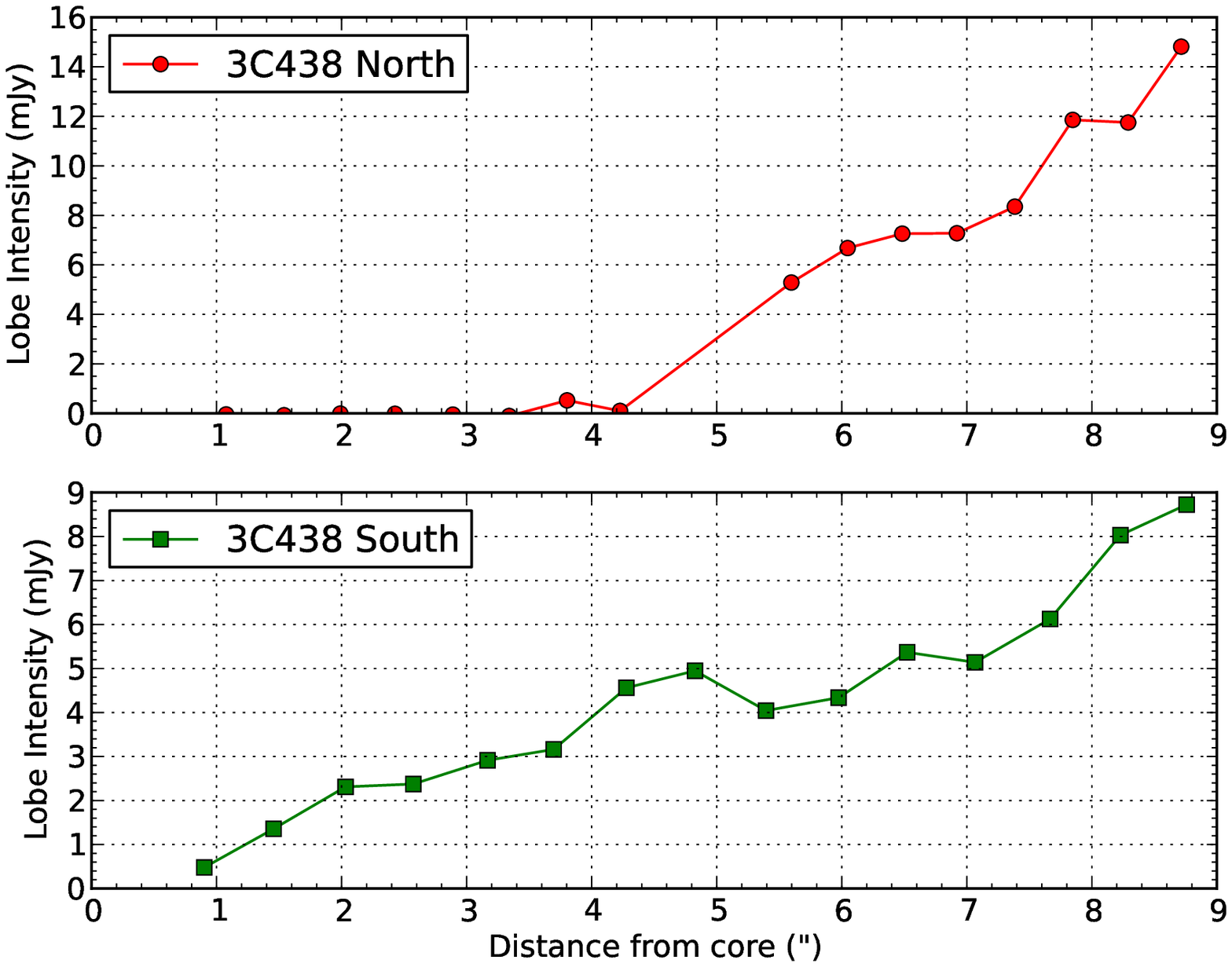}\\

\end{center}
\caption{\label{lobe}Flux density values at different positions for each lobe
(Northeast \& Southwest) compared with other FRII sources \citep{2001ApJ...561..691T}.}
\end{figure}

In Fig. \ref{lobe} we compare our target galaxy to the
FRII galaxy 3C438.
We selected 3C438 for comparison since it is similar in apparent shape compared to J0808.
In both sources we see in general an increasing intensity distribution from the
center towards the edges. Hence the radio structure of 
SDSS~J080800.99+483807.7 is indicative of an edge-brightened FRII source.

\begin{table*}
\begin{center}
\caption{Observed emission lines: wavelengths, redshifts, flux densities, and
line widths}
\label{FWHM}      
\begin{tabular*}{\linewidth}{@{\extracolsep{\fill}}c c c c c c}  
\hline\hline
channel & line & wavelength & redshift & flux & FWHM\\
        &        & [$\AA$]         &     &[$10^{-17}\ \mathrm{erg}\ \mathrm{s}^{-1} \ \mathrm{cm}^{-2}$]& [$\mathrm{km}\ \mathrm{s}^{-1}$]\\                
\hline 
    Blue &  [\ion{O}{ii}] $\lambda$3727   & 4772.1  & 0.2804 & $18.80\pm0.07$    &$445\pm8$\\
    Blue &  H$\gamma$ $\lambda$4340   & 5555.9  & 0.2801 & $4.01\pm0.13$      &$225\pm7$\\
    Red  & H$\beta$ $\lambda$4861   & 6226.5  & 0.2809 & $6.91\pm0.04$      &$333\pm10$\\
    Red  & [\ion{O}{iii}] $\lambda$4959  & 6350.9  & 0.2806 & $4.04\pm0.02$      &$345\pm10$\\
    Red  & [\ion{O}{iii}] $\lambda$5007 & 6412.5  & 0.2807 & $15.03\pm0.02$     &$355\pm10$\\
    Red  & [\ion{He}{i}] $\lambda$5876 & 7523.7 & 0.2804 & $1.51\pm0.04$     &$252\pm9$\\ 
    Red  & [\ion{O}{i}] $\lambda$6300 & 8066.4  & 0.2803 & $2.92\pm0.03$     &$277\pm8$\\
    Red  & [\ion{N}{ii}] $\lambda$6548  & 8386.1 & 0.2807 & $2.65\pm0.03$     &$302\pm8$\\
    Red  & H$\alpha$ $\lambda$6563  & 8405.0 & 0.2806 & $36.02\pm0.03$     &$290\pm8$\\
    Red  & [\ion{N}{ii}] $\lambda$6583 & 8431.3 & 0.2807 & $5.42\pm0.03$     &$261\pm8$\\
    Red  & [\ion{S}{ii}] $\lambda$6716 & 8602.1 & 0.2808 & $6.56\pm0.04$     &$290\pm8$\\
    Red  & [\ion{S}{ii}] $\lambda$6731 & 8621.2 & 0.2808 & $6.25\pm0.04$     &$324\pm8$\\

\hline                                            
\end{tabular*}
\end{center}
\end{table*}

\subsection{Radio loudness}
\label{sec:radio}
Our source is a prominent radio source. The differentiation between
radio loud (RL) and radio quiet (RQ) sources 
primarily reflects the different importance of strong emission from
radio jets and/or lobes \citep[e.g.,][]{2008ApJ...685..801Y}.

\cite{1981A&A....97L...1K} introduced a formalism to derive radio loudness 
by comparing the flux density at 5GHz to the flux density at $4400 \AA$:
\begin{equation}
  R^{*} = \frac{S_{\mathrm{5GHz}}}{S_{4400 \AA}}\ .
\end{equation} 
Currently, there is no flux density measurement at 5~GHz, yet we do have several
flux measurements at different frequencies, as shown in Table~\ref{flux_freq}. From
those, we estimate the spectral index via \citep[cf.][]{2009A&A...494..471O}
\begin{equation}
  \alpha^{v1}_{v2}= \frac{\log\frac{S_{v1}}{S_{v2}}}{\log\frac{v1}{v2}}\ .
\end{equation}
Here $S_{v1}$ and $S_{v2}$ are flux densities at frequencies $\nu_{1}$ and $\nu_{2}$.
From the data at 4.85~GHz and 365~MHz 
in Table~\ref{flux_freq}, we find 
a spectral index of $\alpha=-0.80$. This indicates a dominant contribution to the radio flux from
optically thin synchrotron radiation. Assuming a power law spectrum with
  $S_{v}=\mathrm{const.} \times \nu^{\alpha}$, we find a rest-frame 5~GHz flux density of $S_{\rm 5GHz}=0.076\ \mathrm{Jy}$.
From our optical spectrum, we find a flux density at $4400 \AA$ of 
$S_{4400 \AA}=6.06\times 10^{-6}\ \mathrm{Jy}$.
This results in an estimate of the radio loudness of $R^{*}= 12657$ or $\log R^{*}=4.1$.
We can also compare it to other objects following \cite{2003NewAR..47..593B}. 
To do so, we need to calculate the radio luminosities at 5~GHZ and 178~MHz via
\begin{equation}
    L_{v}= 4\pi \times D_\mathrm{L}^{2} \times S_{v}
\end{equation}
and find:
\begin{eqnarray*}
\log\left(L_{5 \mathrm{GHz}}/\mathrm{erg} \ \mathrm{s}^{-1}\ \mathrm{Hz}^{-1}\right)~~~= 32.27\\
\log\left(L_{5 \mathrm{GHz}}/\mathrm{W}\ \mathrm{Hz}^{-1} \ \mathrm{sr}^{-1}\right)~~~= 25.16\\
\log\left(L_{178 \mathrm{MHz}}/\mathrm{erg} \ \mathrm{s}^{-1}\ \mathrm{Hz}^{-1}\right)= 33.83\\
\log\left(L_{178 \mathrm{MHz}}/\mathrm{W}\ \mathrm{Hz}^{-1} \ \mathrm{sr}^{-1}\right)= 26.33\ .
\end{eqnarray*}

\begin{figure}
\includegraphics*[width=\columnwidth]{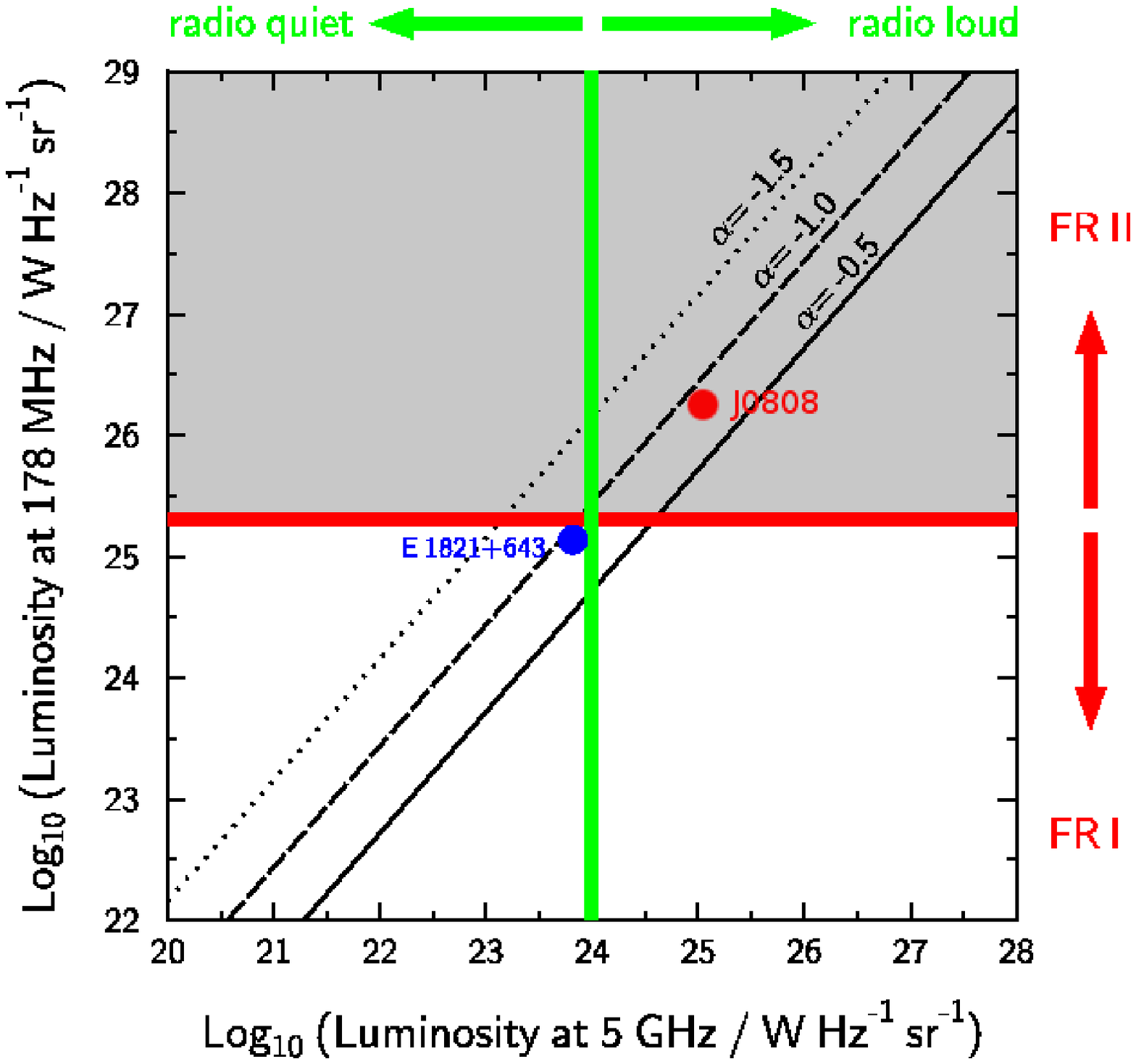}
\caption{Radio loudness scheme \citep{2003NewAR..47..593B}.}
\label{radio_loud}
\end{figure}

In Figure~\ref{radio_loud}, we show the comparison of our RL target with other sources.
This places the J0808 among the RL objects. 
This general result does not change, even if we do the calculation only for the nuclear 
component, which is an order of magnitude weaker than the overall radio luminosity.

\begin{table*}
\begin{center}
\caption{Radio flux density at different frequencies.}   
\label{flux_freq}     
\begin{tabular}{c c c r c}      
\hline\hline
No.   &Observed Passband& Flux             &Luminosity                                &References\\
      &Frequency        & Density          &                                          &          \\
      & (GHz)           &(Jy)              & ($10^{25}\ \mathrm{W}\ \mathrm{Hz}^{-1}$)&          \\
\hline 
    1 & 4.85            & $0.077\pm 0.010$&1.902                                      & 1        \\
    2 & 4.85            & $0.075\pm 0.011$&1.853                                      & 2        \\
    3 & 1.66            & $0.088\pm 0.004$&2.175                                      &this study\\
    4 & 1.4             & $0.162\pm 0.005$&4.003                                      & 3        \\
    5 & 0.365           & $0.563\pm 0.053$&13.913                                     & 4        \\    
\hline                                           
\end{tabular}
\tablebib{
(1)~\citet{1991ApJS...75.1011G}; (2) \citet{1991ApJS...75....1B}; (3) \citet{1998AJ....115.1693C}; (4) \citet{1996AJ....111.1945D}.
}
\end{center}
\end{table*}

\begin{figure}
\includegraphics[width=\columnwidth]{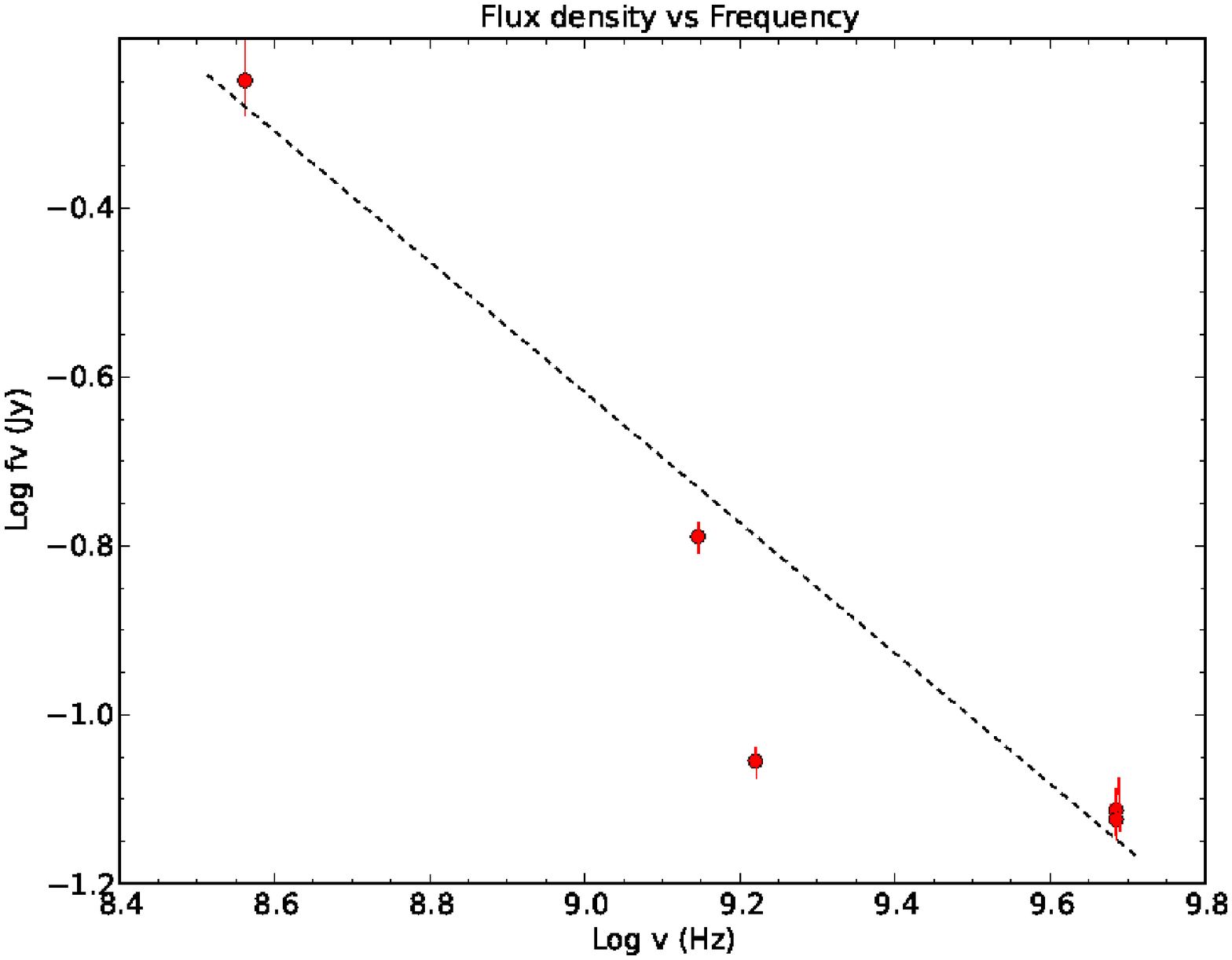}
\caption{Flux density and uncertainty versus frequency of J0808 from 
different surveys \citep{1991ApJS...75.1011G,1991ApJS...75....1B,1998AJ....115.1693C,1996AJ....111.1945D}.}
\label{fl_fr}
\end{figure}

In Figure~\ref{fl_fr} we show the radio spectrum of our target source.
It can be represented as a straight power law. The fact that our flux density
value at 18~cm wavelength lies below the power law fit is most likely due to
resolution effects. It indicates that not all the flux density of the source was
measured on the shortest baselines of our interferometric MERLIN observations.

\subsection{The central black hole of J0808}
\label{sec:bh}
There is a large body of evidence that indicates that most galaxies harbor a
super-massive black hole at their centers \citep [e.g.,][and references
therein]{1998AJ....115.2285M, 
2000ApJ...539L...9F, 2000ApJ...539L..13G, 2001ARA&A..39..309M,
2005SSRv..116..523F, 2013arXiv1304.7762K}.  
Here, we estimate the black hole mass of J0808 via the scaling 
relation between black hole mass and stellar
velocity dispersion of the host bulge component.
Following \cite{2000ApJ...539L...9F}, we used the relation 
\begin{equation}
      \log \frac{M_{\rm BH}}{M_{\odot}} = 8.12 + \log \left(
      \frac{\sigma_*}{200~\mathrm{km}\ \mathrm{s}^{-1}} \right)^{4.24 \pm 0.41}\ .
\end{equation}
Assuming a Gaussian line shape and an instrumental resolution of 150~$\mathrm{km}\ \mathrm{s}^{-1}$, we obtain a value for $\sigma_*$ via the
following expression \citep{2007ApJ...660.1072W,2005ApJ...627..721G}:
\begin{equation}
      \sigma_*=\frac{\sqrt{\left(\mathrm{FWHM}_{[\ion{O}{iii}]}\right)^2-\left(150\ \mathrm{km}\ \mathrm{s}^{-1}\right)^{2}}/2.35}{{1.34}}\ .
\end{equation}

From Table \ref{FWHM}, the FWHM of [\ion{O}{iii}] is $355~\mathrm{km}\
\mathrm{s}^{-1}$. 
Applying this value, we find $\sigma_*= 102.2~\mathrm{km}\ \mathrm{s}^{-1}$ and
a black hole mass of
\begin{equation}
\log \left(\frac{M_\mathrm{BH}}{M_{\odot}} \right)=6.88\pm0.12\ .
\end{equation}

The [\ion{O}{III}] line strength turned out to exhibit large scatter in the scaling
relation since in AGN the [\ion{O}{III}] line emission is often associated with outflows from
the active nucleus \citep{2003ApJ...585..647B}. The [\ion{S}{II}] emission lines
appear to be a better estimator of the underlying stellar velocity dispersion
of the bulge \citep[cf.][]{2007ApJ...667L..33K}.
For comparison, we also determine the black hole mass using the 
[\ion{S}{II}] emission lines with the following equation:
\begin{equation}
      \sigma_*=\frac{\sqrt{\left(\mathrm{FWHM}_{[\ion{S}{II}]}\right)^{2}-\left(150\ \mathrm{km}\ \mathrm{s}^{-1}\right)^2}}{{2.35}}\ .
\end{equation}

For the two [\ion{S}{II}] lines in the spectrum (see Table~\ref{FWHM}), we
obtain an average value of
$\mathrm{FWHM}\left([\ion{S}{II}]\right)_\mathrm{avg}=307\ \mathrm{km}\
\mathrm{s}^{-1}$, resulting in a value of $\sigma_*=114.0\ \mathrm{km}\
\mathrm{s}^{-1}$. Following \cite{2007ApJ...667L..33K} we find the 
estimation of black hole mass is
\begin{equation}
\log \left(\frac{M_{\mathrm{BH}}}{M_{\odot}}\right)=6.9\pm0.15\ .
\end{equation}
Both methods result in a consistent estimate of the black hole mass of
about $10^7M_\odot$.

The Eddington ratio $\eta$ is a measure of the accretion efficiency because it relates 
the observed AGN radiation output to that 
produced by maximum spherical accretion with isotropic radiation.
The Eddington ratio $\eta=L_\mathrm{bol}/L_\mathrm{edd}$ is the ratio between
the bolometric luminosity $L_\mathrm{bol}$ and the 
Eddington luminosity $L_\mathrm{edd}$. 
The bolometric luminosity is usually estimated from single-band
measurements. Depending on the particular spectral energy distribution, there is 
considerable uncertainty in the bolometric correction for individual sources.
Here, we estimate the bolometric luminosity, following the prescription from
\cite{2004ApJ...601..676V}, using the rest-frame 5100$\AA$ luminosity density
\begin{equation}
\label{bol}
    L_\mathrm{bol}\approx 9.47\times \lambda L_\lambda(5100\AA)\ \mathrm{erg~s}^{-1}.
\end{equation}

\begin{figure}
\includegraphics[width=\columnwidth]{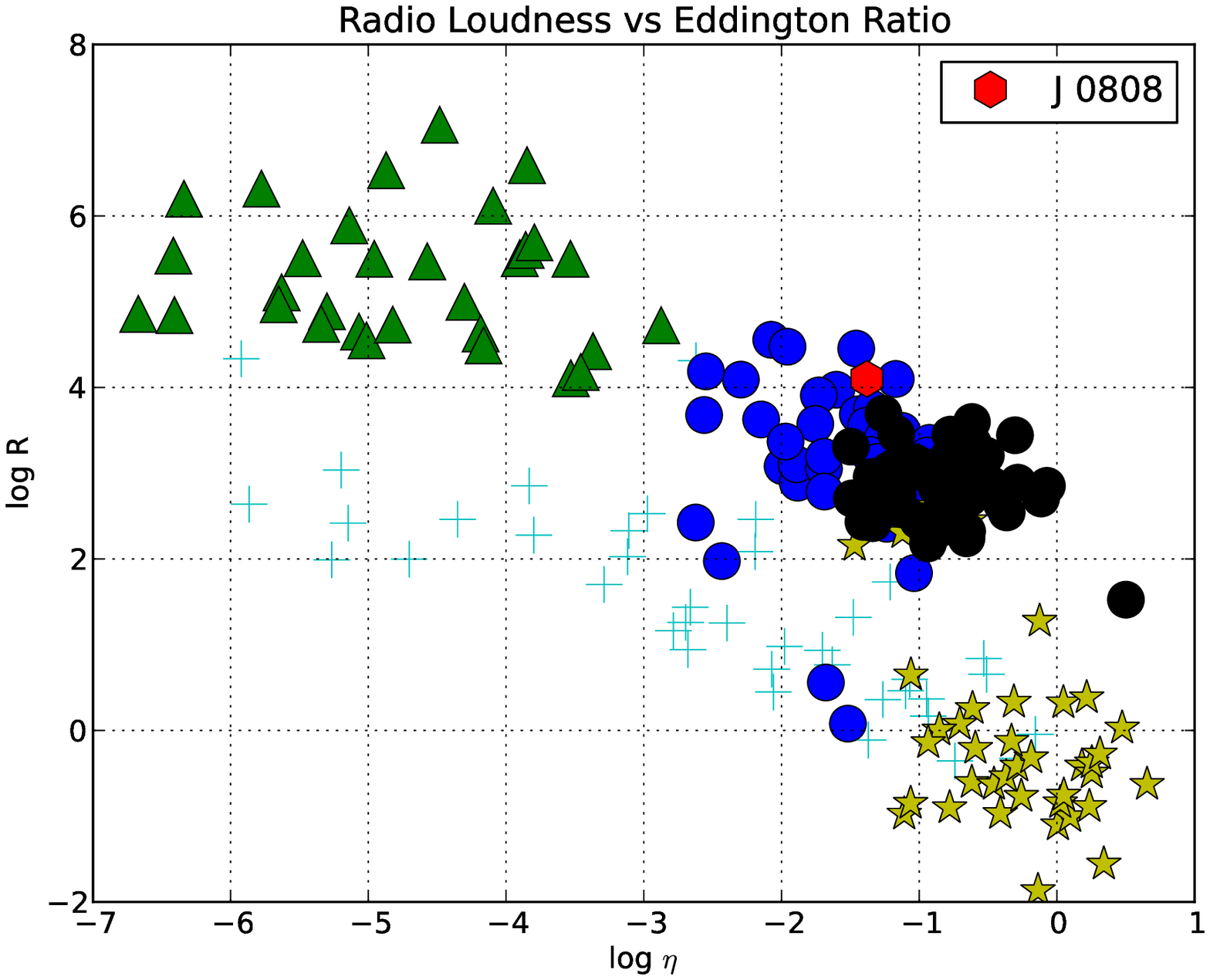}
\caption{Radio loudness (R) versus Eddington ratio $\eta$. Our target source is represented by a red hexagon, 
BLRGs are marked by the blue circles, radio loud quasars by the Black circles, 
Seyfert galaxies and LINERs by the crosses, FR I radio galaxies by the green triangles, and
PG quasars by the yellow stars \citep{2007ApJ...658..815S}.}
\label{RL}
\end{figure}

\begin{figure}
\includegraphics[width=\columnwidth]{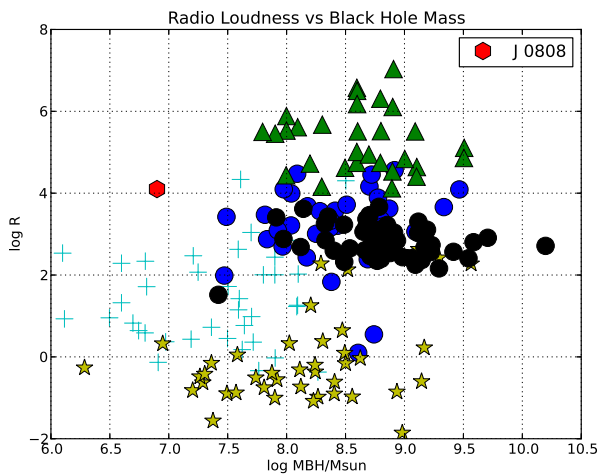}
\caption{Radio loudness versus black hole mass.
See caption of Fig.~\ref{RL}.}
\label{BH}
\end{figure}

The rest-frame 5100$\AA$ luminosity density is estimated by fitting a
linear combination of two simple starburst models, the AGN power law, and an
extinction component to the observed spectrum \citep[cf.][]{2012A&A...543A..57Z}. 
We find a 5100$\AA$ continuum luminosity of 
$\lambda L_\lambda(5100\AA)=7.4\times 10^{42}\ \mathrm{erg}\ \mathrm{s}^{-1}$.
From Eq \ref{bol}, the bolometric luminosity is 
$L_\mathrm{bol} \approx7.0\times 10^{43}\ \mathrm{erg}\ \mathrm{s}^{-1}$, and 
with a black hole mass estimate of $\log\left(M_{BH}/M_{\odot}\right)\thicksim6.9$, 
the Eddington luminosity\footnote{The Eddington luminosity is the luminosity radiated at the Eddington limit 
and is calculated as $L_\mathrm{edd}\approx 1.26\times 10^{38} \left(\frac{M_\mathrm{BH}}{M_{\odot}}\right) \mathrm{erg}\ \mathrm{s}^{-1}$.}
is $L_\mathrm{Edd}\approx 1.6 \times 10^{45}
\ \mathrm{erg}\ \mathrm{s}^{-1}$. This results in a value for the Eddington ratio of $\log~\eta=-1.39$.
In Figure~\ref{RL} and Figure~\ref{BH} we compare our values 
with published results.
As shown in Fig. \ref{RL}, in $\log \eta - \log R$ space, the source 
J0808 lies in the region shared by broad emission-line radio-galaxies (BLRG) 
and RL quasars. The same is found in Fig. \ref{BH} ($\log M_\mathrm{BH} - \log R$). 
However, the mass of the super-massive black hole appears to  
be lower than that of the bulk of radio galaxies considered in the plot.

\begin{table*}
\begin{center}
\caption{The rest radio flux densities and luminosities for source components of the J0808 radio counterpart at 18 cm.}
\label{phy_meas}      
\begin{tabular}{c l l l c l l c l l l }
\hline\hline       
Position&\multicolumn{2}{c}{Brightest Point}& \multicolumn{2}{c}{Center}&\multicolumn{2}{c}{Total Area}\\
\hline
        &Flux &Luminosity    &Flux&Luminosity      &Flux&Luminosity\\
        &$(10^{-3}\ \mathrm{Jy}$&$(10^{23}\ \mathrm{W}\ \mathrm{Hz}^{-1}$&$(10^{-3}\ \mathrm{Jy}$&$(10^{23}\ \mathrm{W}\ \mathrm{Hz}^{-1}$&($10^{-2}\ \mathrm{Jy})$&$(10^{24}\ \mathrm{W}\ \mathrm{Hz}^{-1})$\\
        &per beam)      & per beam)       & per beam)    & per beam) & & \\
\hline
Southeast Lobe &1.28&3.17&0.81&2.01&3.71&9.17\\
Center Region &0.27&0.67&--  & -- &0.72&1.79\\
Northwest Lobe &1.41&3.49&0.45&1.12&4.69&11.59\\
\hline                  
\end{tabular}         
\end{center}
\end{table*}

\subsection{Star formation}
\label{sec:star_form}
The hosts of radio galaxies may be characterized by strong star-formation activity.
However, this is a matter of intense debate as interactions between the jet and the 
host ISM may also lead to suppression of star formation, as discussed in, e.g., \cite{2010A&A...521A..65N}.
Radio galaxies with strong nuclear emission lines have star-formation activity that is
a factor of 3-4 higher than in weak-line sources \citep{2013MNRAS.429.2407H}.
The line emission of our target source and the fact that it is RL suggests 
that it has a substantial star-formation rate. 

To estimate the star-formation rate from optical data, we use the 
luminosity of the emission line [\ion{O}{ii}] $\lambda$3728
\citep{2003ApJ...599..971H,1998ARA&A..36..189K,2005ApJ...629..680H}: 
\begin{equation}
{\mathrm{SFR}_{\mathrm [\ion{O}{ii}]}}\,[M_{\odot}\,\mathrm{yr}^{-1}] =
   \frac{L_{\mathrm [\ion{O}{ii}]}}{2.97\times10^{33}\,{\mathrm W}}\ .
\end{equation}

Here we correct the reddening of the [\ion{O}{ii}] using theoretical Balmer decrement $\mathrm{H}\alpha / \mathrm{H} \beta = 2.86$ and assuming 
case B recombination for a region with electron density of $10^{4}\ \mathrm{cm}^{-3}$ and temperature $T=10^{4}\ \mathrm{K}$ \citep{1989agna.book.....O}. 
According to the extinction law of \cite{1994ApJ...429..582C},  
we obtain the luminosity of [\ion{O}{ii}] $\lambda$3728 of $5.82\times 10^{34} \mathrm{W}$, resulting in a star-formation 
rate of $\mathrm{SFR}= 19.59 M_\odot\ \mathrm{yr}^{-1}$.
In addition, we can make use of the WISE infrared survey data on J0808. There 
the source has been measured at $3.4$, $4.6$ and $12\mu$m. 
From a sample of dusty galaxies, we randomly picked galaxies from the IRAS point source catalog \citep{1985cgqo.book.....L} and
find that the median flux density ratio between 12$\mu$m and 60$\mu$m wavelength lies at about 20 with an uncertainty of
about 10. From \cite{1986ApJ...311L..33H}, we know that the 
ratio between the flux densities $S_{60}$ and $S_{100}$ at 
100$\mu$m and 60$\mu$m wavelength is of the order of $2\pm1$.
Hence, we can extrapolate from the WISE detection of J0808 at 12$\mu$m to obtain an
estimate of the far-infrared (FIR) luminosity of 
$L_\mathrm{FIR}=23.96\times 10^5 \left(2.258~S_{60}+S_{100}\right)D_L[\mathrm{Mpc}]^2=(2.2\pm1.1) \times 10^{10} L_{\odot}$.
This results in a star-formation rate of OBA stars of
$\nu_\mathrm{OBA} =2.1 \times 10^{-10} \times  L_\mathrm{FIR}=5 \pm 4\ M_{\odot}\,\mathrm{yr}^{-1}$
\citep{1986ApJ...311...98T}.
To measure the star-formation rate from radio data, we apply the formalism published 
by \cite{2003ApJ...586..794B}, which is related to the 20 cm luminosity density
\begin{equation}
	\left(\frac{\mathrm{SFR}}{M_\odot \mathrm{yr}^{-1}}\right) = 
5.52\times 10^{-22}\left(\frac{L_\mathrm{20cm}}{\mathrm{W\,Hz}^{-1}}\right).
\label{eq:radioSFR}
\end{equation}
From the radio spectrum discussed earlier, we obtain the luminosity at 20 cm
and find $L_{20\ \mathrm{cm}}$=$6\times 10^{23}\ \mathrm{W}\ \mathrm{Hz}^{-1}$. Interpreting this radio luminosity in the
context of star formation, this implies that only 1\% of the total nuclear radio 
luminosity is required to explain the expected contribution from the above derived 
star-formation rate. The bulk of the emission may be due to the nuclear jet or the 
portion of the jet that flows back along the envelope of the jet onto the host.

The radio emission can also be due to contributions from supernova explosions.
The rate of supernova explosions is a measure of the high-mass star formation.
Following \cite{1990ApJ...357...97C}, we calculate the supernova rate of our target galaxy 
from its central radio flux density via
\begin{equation}
   \left(\frac{\mathrm{SNR}}{\mathrm{yr}^{-1}}\right)
   \approx 7.7\times 10^{-24}\left(\frac{\nu}{\mathrm{GHz}}\right)^\alpha \left(\frac{L_\mathrm{NT}}{\mathrm{W}\mathrm{Hz}^{-1}}\right)\ .
\end{equation}
Here, SNR is the supernova rate per year,
$\nu$ is the frequency,
$L_\mathrm{NT}$ the nonthermal radio luminosity, and 
$\alpha$ the radio spectral index.

From our MERLIN observations at 18 cm, we find 
a central luminosity at 18~cm of
${L_\mathrm{18 cm}=0.67\times10^{23}\ \mathrm{W}\ \mathrm{Hz}^{-1}}$ from table~\ref{phy_meas}. 
For an assumed nuclear spectral index of $\alpha=-0.60$, we calculate the supernova rate at the center
as $\mathrm{SNR}\approx0.4\ \mathrm{yr}^{-1}$. 
This is consistent with the OBA star-formation rate obtained above from the 
extrapolation of the WISE 12$\mu$m measurements. 
However, given that the nuclear radio emission may contain a significant 
amount of nonthermal radiation from the jet 
and taking into account that the nuclear spectral index could also be steeper than $\alpha=-0.60$, 
we consider $\mathrm{SNR}$ as an upper limit.
Following \cite{2001AJ....121..128H} and \cite{2006AJ....131..701L}, the host is considered to be 
a Seyfert galaxy if the  supernova rate is $\lesssim 1\ \mathrm{yr}^{-1}$.
Starburst galaxies have supernova rates of $\mathrm{SNR}\lesssim 10\ \mathrm{yr}^{-1}$. 

\section{Discussion and Conclusions}
\label{sec:discussion}
Through MERLIN radio observations we have shown that the 
source J0808 is a compact double lobed radio source 
at a redshift of $z=0.2805\pm 0.0003$; it shows a radio structure reminiscent of those 
that may present significant back-flow of material along the jet or outflow into the host 
\citep{2013ApJ...763L..18W,2013ApJ...772..112S}.
The host of the radio source has been detected in the optical and infrared emission 
in the SDSS and WISE surveys.
The optical spectrum as well as the optical and infrared images of the 
host suggest that it is a compact early-type host galaxy. 
Our LBT MODS1R spectra, an estimate of the FIR luminosity, and the radio flux at the position of the 
host reveal that the host shows indications of strong star-forming activity.  
The widths of forbidden lines indicate a black hole mass of $\sim10^{6.9}$\solm.

The detection of the binarity of sources also clearly depends on the size of the telescope or
interferometer used. In order to make sure that the location of  J0808
at the lower edge of the distribution shown in Fig.5 is indeed special, we compare the properties
of small and large radio double sources that, in fact, form a continuum in separations.
\cite{1997AJ....113..148O} plot power versus projected largest linear size for 
the complete samples of gigahertz-peaked spectrum (GPS) and compact-steep spectrum (CSS) sources
\citep{1998A&AS..131..303S,2009AN....330..303F,1995A&A...302..317F,1985A&A...143..292F} and the 3CR 
\citep{1983MNRAS.204..151L} for the redshift range $0.2 \le z \le 1.0$ and find that there is a good overlap between the samples.

The radio source evolution can be constrained by plotting the number of sources as a function of size in the 
power versus linear size plane. \cite{1997AJ....113..148O} find that the number of sources 
is roughly constant per linear size bin for sources less than about 10~kpc.
For the larger sources, the number increases with increasing 
source size.
This can be interpreted as an indication for a qualitatively different evolution of the small sources 
compared to the large ones.
In fact, \cite{1998PASP..110..493O} points out that the GPS sources are entirely contained within the
extent of the narrow-line region (less than about 1 kpc), while the CSS 
sources are contained entirely within the host galaxy (less than about 15 kpc).
An explanation for this effect may be that the small sources are still embedded in the ISM of the host 
while the large ones can expand more freely.
This shows that the source we are presenting here is indeed representative for the 
smaller (around 10~kpc size) of the large double sources discussed by \cite{1997AJ....113..148O}.

Although the hosts of double radio sources in general do not have significant amounts of 
molecular gas, they do show signatures of star formation.
An exception, on the one hand, are starburst radio galaxies that comprise 15-25 per cent of all powerful
extragalactic radio sources \citep{2011MNRAS.412..960T}.
On the other hand, it appears that the interaction with the radio jet
quenches star formation and that hosts of radio galaxies seem to be inefficiently
forming stars \citep[e.g.,][]{2010A&A...521A..65N}.
\cite{2005AJ....129..610O} find no abundant molecular gas reservoirs in GPS radio sources with upper limits of 
10$^9$ to a few times 10$^{10}$ \solm.

Despite the apparent lack of large amounts of molecular gas,
\cite{2011A&A...528A.110F} and \cite{2008A&A...477..491L} find that most of the hosts of CSS 
radio binaries show an excess of ultraviolet (UV) radiation compared to the spectra of local RQ ellipticals. 
This UV excess may be due to an active nucleus or to a young stellar population, both of which may be 
triggered by the same event or influence each other.
Similarly, we also find for J0808 diagnostic line ratios that clearly indicate the 
presence of star formation in the host galaxy.
While a merger event may have caused both the ongoing jet activity, we have, however, 
no indication that the host has been affected by a recent merger event from the 
available imaging information. Hence the case of J0808 also allows us to follow a different interpretation: 
This source has an extended nuclear radio structure suggesting 
an interaction of the host ISM with the back-flowing material. 
Thus for J0808 the star formation may be triggered by the 
back-flow along the jet and its interaction with the AGN host.

\begin{acknowledgements}
The authors kindly thank the entire MERLIN staff for excellent support. 
We also thank the anonymous referee for the useful comments and
suggestions that helped to improve the paper. 
Y.E. Rashed is supported by the German Academic Exchange Servis (DAAD)
and by the Iraqi ministry of higher education and scientific research.
G.B. is member of the Bonn-Cologne Graduate School of Physics and Astronomy 
(BCGS) and acknowledges support from the Konrad-Adenauer-Stiftung (KAS).
M.V-S. thanks the funding from the
European Union Seventh Framework Programme (FP7/2007-2013)
under grant agreement No.312789.
M.V. is member of the International Max-Planck Research School (IMPRS)
for Astronomy and Astrophysics at the Universities of Bonn and Cologne.
The work is also supported in part by the Deutsche Forschungsgemeinschaft (DFG)
via grant SFB 956.
We had fruitful discussions with members of the European Union
funded COST Action MP0905: Black Holes in a violent Universe and the
COST Action MP1104: Polarization as a tool to study the Solar System and beyond.
This work has benefited from research funding from the European Community's sixth
Framework Programme under RadioNet R113CT 2003 5058187.
Radio observations were made with MERLIN, a National Facility operated by the University of Manchester at
Jodrell Bank Observatory on behalf of STFC.
The spectroscopic observations reported here were obtained at the LBT
Observatory, a joint facility of the Smithsonian Institution and the University of Arizona.
LBT observations were obtained as part of the Rat Deutscher Sternwarten
guaranteed time on Steward Observatory facilities through
the LBTB coorperation.
This paper uses data taken with the MODS spectrographs built with funding from
NSF grant AST-9987045 and the NSF Telescope System Instrumentation Program (TSIP),
with additional funds from the Ohio Board of Regents and the Ohio State University Office of Research. 
The Sloan Digital Sky Survey is a joint
project of the University of Chicago, Fermilab, the Institute for Advanced
Study, the Japan Participation Group, Johns Hopkins University, the Max-Planck-Institute 
for Astronomy, the Max-Planck-Institute for Astrophysics, New Mexico
State University, Princeton University, the United States Naval Observatory, and
the University of Washington. Apache Point Observatory, site of the SDSS, is
operated by the Astrophysical Research C
onsortium. Funding for the project has
been provided by the Alfred P. Sloan Foundation, the SDSS member institutions,
NASA, the NSF, the Department of Energy, the Japanese Monbukagakusho, and
the Max-Planck Society. The SDSS Web site is
http://www.sdss.org.
Y.E.R. wants to thank all group members at I. Institute of Physics
as well as Dr. Peter A. Schuller for fruitful discussions and help in the preparation of this publication.
\end{acknowledgements} 

\bibliographystyle{bibtex/aa}
\bibliography{Yasir_E_Rashed} 

\end{document}